\documentclass[journal]{IEEEtran}
\usepackage{cite}

\usepackage{color,soul}
\usepackage{gensymb}
\usepackage{dsfont}

\ifCLASSINFOpdf
  \usepackage[pdftex]{graphicx}
  \graphicspath{{../pdf/}{../jpeg/}}
  \DeclareGraphicsExtensions{.pdf,.jpeg,.png}
\else
  \usepackage[dvips]{graphicx}
  \graphicspath{{../eps/}}
  \DeclareGraphicsExtensions{.eps}
\fi
\usepackage[cmex10]{amsmath}
\usepackage{amssymb}
\DeclareMathOperator*{\argmaxA}{arg\,max}

\usepackage{euscript}

\usepackage{multirow}
\usepackage{algorithm}
\usepackage[noend]{algpseudocode}
\usepackage{diagbox}
\usepackage{physics}
\usepackage{tikz}

\ifCLASSOPTIONcompsoc
  \usepackage[caption=false,font=normalsize,labelfont=sf,textfont=sf]{subfig}
\else
  \usepackage[caption=false,font=footnotesize]{subfig}
\fi
\usepackage{url}

\begin{document}
%
\title{A Multi-Timescale Data-Driven Approach to Enhance Distribution System Observability}
%
%
%

\author{Yuxuan~Yuan,~\IEEEmembership{Student Member,~IEEE,}
	Kaveh~Dehghanpour,~\IEEEmembership{Member,~IEEE,}
	Fankun~Bu,~\IEEEmembership{Student Member,~IEEE,}
	and Zhaoyu~Wang,~\IEEEmembership{Member,~IEEE}
\thanks{This work is supported by the U.S. Department of Energy Office of Electricity Delivery and Energy Reliability under DE-OE0000875.

 Y. Yuan, K. Dehghanpour, F. Bu, and Z. Wang are with the Department of
Electrical and Computer Engineering, Iowa State University, Ames,
IA 50011 USA (e-mail: yuanyx@iastate.edu; wzy@iastate.edu).
 }
}
%
%

\markboth{Submitted to IEEE for possible publication. Copyright may be transferred without notice}%
{Shell \MakeLowercase{\textit{et al.}}: Bare Demo of IEEEtran.cls for Journals}
%



\maketitle

\begin{abstract}
This paper presents a novel data-driven method that determines the daily consumption patterns of customers without smart meters (SMs) to enhance the observability of distribution systems. Using the proposed method, the daily consumption of unobserved customers is extracted from their monthly billing data based on three machine learning models: first, a spectral clustering (SC) algorithm is used to infer the typical daily load profiles of customers with SMs. Each typical daily load behavior represents a distinct class of customer behavior. In the second module, a multi-timescale learning (MTSL) model is trained to estimate the hourly consumption using monthly energy data for the customers of each class. The third stage leverages a recursive Bayesian learning (RBL) method and branch current state estimation (BCSE) residuals to estimate the daily load profiles of unobserved customers without SMs. The proposed data-driven method has been tested and verified using real utility data.    
\end{abstract}

\begin{IEEEkeywords}
	Observability, spectral clustering, machine learning, distribution system state estimation.
\end{IEEEkeywords}

%
\IEEEpeerreviewmaketitle

\section{Introduction}
Advanced Metering Infrastructure (AMI) enables utilities to perform energy consumption measurement, demand-side control, tampering detection, and voltage monitoring \cite{amireport}. The core element of AMI is smart meters (SMs). Compared to conventional electromechanical meters that simply record the monthly energy consumption data, SMs record the real-time load consumption of customers. Recently, a rapid growth of SMs has been observed in distribution systems. According to statistical data provided by the U.S. Energy Information Administration (EIA), the nationwide number of SMs was estimated to be 70.8 millions in 2016 with an annual growth of 6 million devices from the previous year \cite{eiasmart}. Nonetheless, due to financial limitations and cyber-security issues, the number of SMs in many distribution networks is still limited. Hence, many utilities still rely on traditional monthly consumption data to obtain load behaviors. This lack of knowledge of real-time load behaviors inhibits effective monitoring and control of the system. One approach for solving this problem is to widely install SMs, which is cost prohibitive. As an alternative solution, we will design data-driven real-time load estimation techniques for inferring customers' behaviors \cite{OCS2009}. 

In recent years, several papers have focused on load estimation, including missing data reconstruction, communication delay compensation, and unobserved customer behavior inference. The previous works in this area can be classified into two categories based on the temporal granularity of customer datasets used for model development: \textit{Class I}: A number of articles use data with at least hourly resolution for training load estimation methods \cite{kmeans2017, Yli2013, Bstep2014, DG2005, EM2012}. In \cite{kmeans2017}, a K-means-based load estimation approach is proposed to estimate the missing measurements by using historical half-hourly energy consumption data. In \cite{Yli2013}, a truncated Fourier series representation and cluster analysis are utilized to estimate a hybrid model of consumer load during summers. In \cite{Bstep2014}, several linear Gaussian load profiling techniques are employed to capture customer behaviour using SM data analysis. In \cite{DG2005}, in addition to SM data, the context information of customers, such as operation time during the weekends and economic codes, are leveraged to allocate the respective load profiles among particular groups, utilizing a probabilistic neural network (PNN)-based approach. In \cite{EM2012}, power flow simulation data with half-hourly temporal resolution is exploited to obtain load estimation using Artificial Neural Networks (ANN). \textit{Class II}: Instead of using data with high temporal resolution, a number of papers estimate the hourly customer energy consumption by converting the monthly billing data into daily load profiles \cite{YR2018, JA2000, DT2015}. In \cite{DT2015}, hourly load estimation is performed using uniform energy allocation, where the mean and variance of estimated load is adjusted in real-time utilizing supervisory control and data acquisition (SCADA) devices. In \cite{YR2018}, typical load profiles are assigned to the unobserved customers by comparing average daily consumption values with the daily energy levels of the representative load profile obtained from observed customers. The pseudo load profiles of unobserved customers are scaled by multiplying the estimated average consumption with the corresponding load pattern. Based on the monthly energy level, the daily load profile of unobserved customer can be obtained using representative curves from statistical analysis of residential, commercial, and industrial consumers' historical data \cite{JA2000}.

While previous works provide valuable results, many questions remain open with respect to the real-time load estimation in distribution systems. For example, accurate performance of Class I models depends on high penetration of real-time measurement units and availability of a sizable data history, which renders their practical implementation costly. On the other hand, Class II methods are generally based on the simplified assumption that the total daily energy consumption for each customer remains almost constant during a month. This assumption reduces the estimation accuracy. While in \cite{YR2018} a separation between weekday and weekend consumption data was introduced to alleviate this problem, this approach falls short of distinguishing load behavior in different individual days.

In order to address these shortcomings, in this paper, a spectral clustering (SC)-based multi-timescale learning (MTSL) framework is proposed to estimate hourly load consumption for customers without SMs, using monthly billing data. Unlike previous Class II methods, the proposed model is able to distinguish the daily customer behaviors. To achieve this, three stages are included in the load estimation framework: 1) Typical daily load profiles are classified and stored in a databank using a SC algorithm trained by the AMI dataset of \textit{observed customers} (i.e., customers with SMs) \cite{LZM2004}. 2) For each class of typical load behavior, a multi-layer MTSL model is developed, which can decompose the monthly consumption into different timescale components, such as weekly, daily, and hourly consumption. At each layer, a series of machine learning models are used to allocate energy consumption at slower timescale among faster timescale consumption variables. 3) Due to the absence of real-time data for unobserved customers without SMs, a branch current state estimation (BCSE)-aided method is proposed to identify their underlying typical daily consumption \cite{MEB1995}. The residuals of BCSE are used to calculate the probability of all classes using a recursive Bayesian learning (RBL) approach \cite{RS2010}. The class with the highest probability is selected as the underlying typical load behavior for the unobserved customer. While this method is trained using SM data from observed distribution systems, it can be employed to estimate the hourly load data for a fully unobservable network without SMs. The proposed method has been tested using real utility data and compared with existing methods in the literature.    

The rest of this paper is constructed as follows: Section \ref{overall} introduces the proposed observability enhancement framework. In Section \ref{cluster}, a SC algorithm
is utilized to build the consumption pattern bank for different types of customers. In Section \ref{learning}, the MTSL method is presented. Section \ref{identify} formulates the BCSE-aided pattern identification approach. The numerical results are analyzed in Section \ref{result}. Section \ref{conclusion} concludes the paper with major findings.

\section{Introduction to Real Data and Proposed Observability Enhancement Framework}\label{overall}
\begin{figure}[htbp]
	\centering
	\includegraphics[width=3.5in]{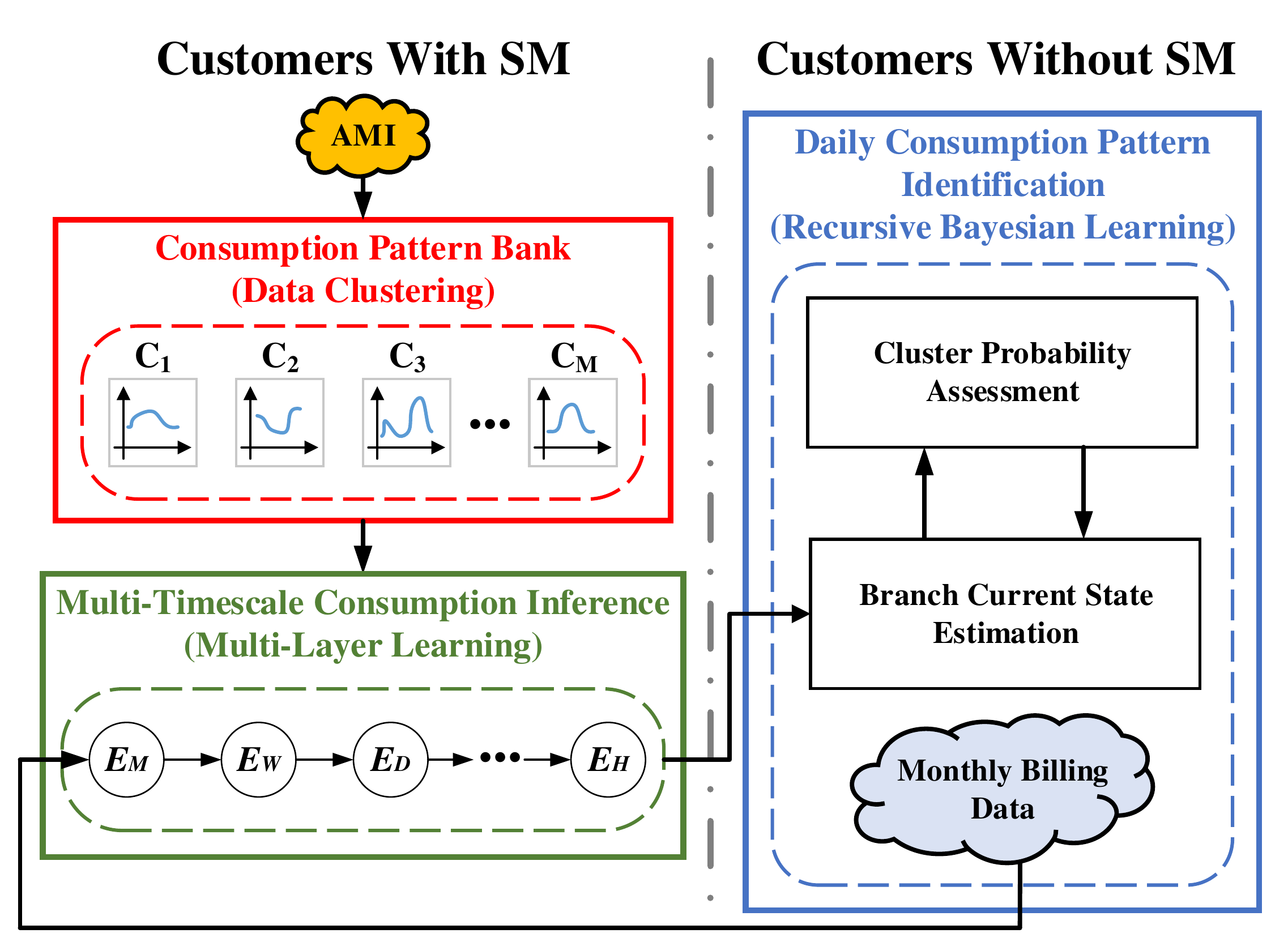}
	\caption{Proposed observability enhancement framework.}
	\label{fig:main}
\end{figure}

\subsection{AMI Data Description}
The available AMI data history contains several U.S. mid-west utilities' hourly energy consumption data (kWh) for over 6000 customers. The data ranges from January 2015 to May 2018. While a few industrial consumers are included in the dataset, over 95\% of customers are residential and commercial loads. The hourly data was initially processed to remove missing data caused by communication error. Then, the AMI dataset was divided into six separate subsets where each subset corresponds to weekday or weekend load profiles of residential, commercial and industrial customers.

\subsection{Proposed Observability Enhancement Framework}
The objective of this paper is to design a load estimation approach for fully or partially unobservable networks to avoid overmuch assumptions in the location/type of measurement units and availability of context information. Given that monthly billing data of consumers is generally available in all distribution systems, the data resource required for training the proposed load estimation approach consists of unobserved customers' monthly billing data and a limited number of AMI data from other observed networks. Extra available context information can also be added to improve the performance of the model but is not required. Different stages of the proposed observability enhancement framework are presented in Fig. \ref{fig:main}.
\begin{itemize}
\item \textbf{Stage I - Consumption Pattern Bank:}
Based on the six data subsets defined above, a SC algorithm is used to detect similarities in the diverse daily load profiles and define customer classes accordingly. As shown in Fig. \ref{fig:main}, the results of clustering, $\{C_1,C_2,...,C_M\}$, are stored in the specific consumption pattern bank according to the customer type, with each cluster representing a typical daily load profile. The pattern bank clustering results are stored and employed for the development of machine learning models (detailed in Section \ref{cluster}).

\item \textbf{Stage II - Multi-Timescale Consumption Inference:}
A separate multi-layer MTSL model is trained for each class of customers using SM data of observed customers to convert the monthly billing data to hourly load values. In each MTSL model, machine learning algorithms are developed based on various pre-determined timescales. The customer consumption at these timescales are defined as monthly consumption $E_M$, weekly consumption $E_W$, daily consumption $E_D$, and hourly consumption $E_H$. The monthly data is regarded as the input for the first layer of the model and the hourly consumption variables appear in the output of the final layer. After the individual MTSL model of different classes are developed, the hourly estimation of unobserved customers are inferred by these models (detailed in Section \ref{learning}).

\item \textbf{Stage III - Consumption Pattern Identification:}
In practice, the real hourly load of unobserved customers are unavailable \textit{a priori} to determine the homologous daily load patterns. Hence, to assign a class from the daily pattern databank (Stage I) to unobserved customers, a BCSE-aided RBL method is proposed to identify these customers' underlying daily load profiles. Different daily profiles and their respective MTSL models are used for running BCSE over the target network for a period of time. The measurement residuals for each daily pattern are observed and utilized to make a connection between unobserved customers and their correct daily consumption patterns. Based on the observed residuals, a RBL method is employed to recursively assign a probability value to each typical daily consumption pattern for each unobserved customer. Then, the model with the highest probability is identified as the ``correct'' daily profile. The MTSL corresponding to the identified class for an unobserved customer is used to generate hourly pseudo measurements for that customer providing the redundancy to enhance the system observability (more details in Section \ref{identify}).

\end{itemize}

\section{Proposed Clustering Algorithm}\label{cluster}

With the advent of AMI systems, typical daily load profile classification can be performed using different clustering algorithms, such as K-means, self-organizing maps, and hierarchical clustering \cite{evacluster2010}. In this paper, a graph theory-based clustering technique known as SC is utilized to distinguish the typical load profiles of observed customers and to create the typical consumption pattern bank. SC algorithm employs eigenvectors of graph matrices for data reconstruction, while using automatic neighbor detection to avoid error from manual parameter selection \cite{LZM2004}. SC treats the data clustering as a graph partitioning problem without making any assumption on the data distribution \cite{GJ2007}. That means SC outperforms traditional clustering techniques when tested on complex and unknown customer load shapes \cite{sc2017}. In this paper, the main steps of SC are listed as follows:

\begin{itemize}
\item \textbf{Step I:}
As a graph theoretic clustering approach, SC algorithm transforms AMI dataset into a similarity graph $G=(V,E)$, which consists of a set of vertices $V$ and a set of edges $E$ connecting different vertices. For our problem, average daily load profile of an observed customer is defined as a vertex $V \in\mathbb{R}^{24}$, where two vertices are connected if the corresponding pair-wise similarity is non-zero. In this paper, a technique is utilized for constructing fully-connected graphs, in which vertex $V_i$ is connected to all vertices that have positive similarity with $V_i$. The goal of similarity graph is to model local neighborhood relations between data points. The value of similarity relies on a scaling parameter $\alpha$ that controls how rapidly the similarity weights, $W_{ij}$, fall off with the distance between vertices. Note that the \textit{distance} between vertices $a$ and $b$ is defined as $||a-b||$ \cite{NMW2002}. 
\begin{figure}[htbp]
\centering
\includegraphics[width=3.25in]{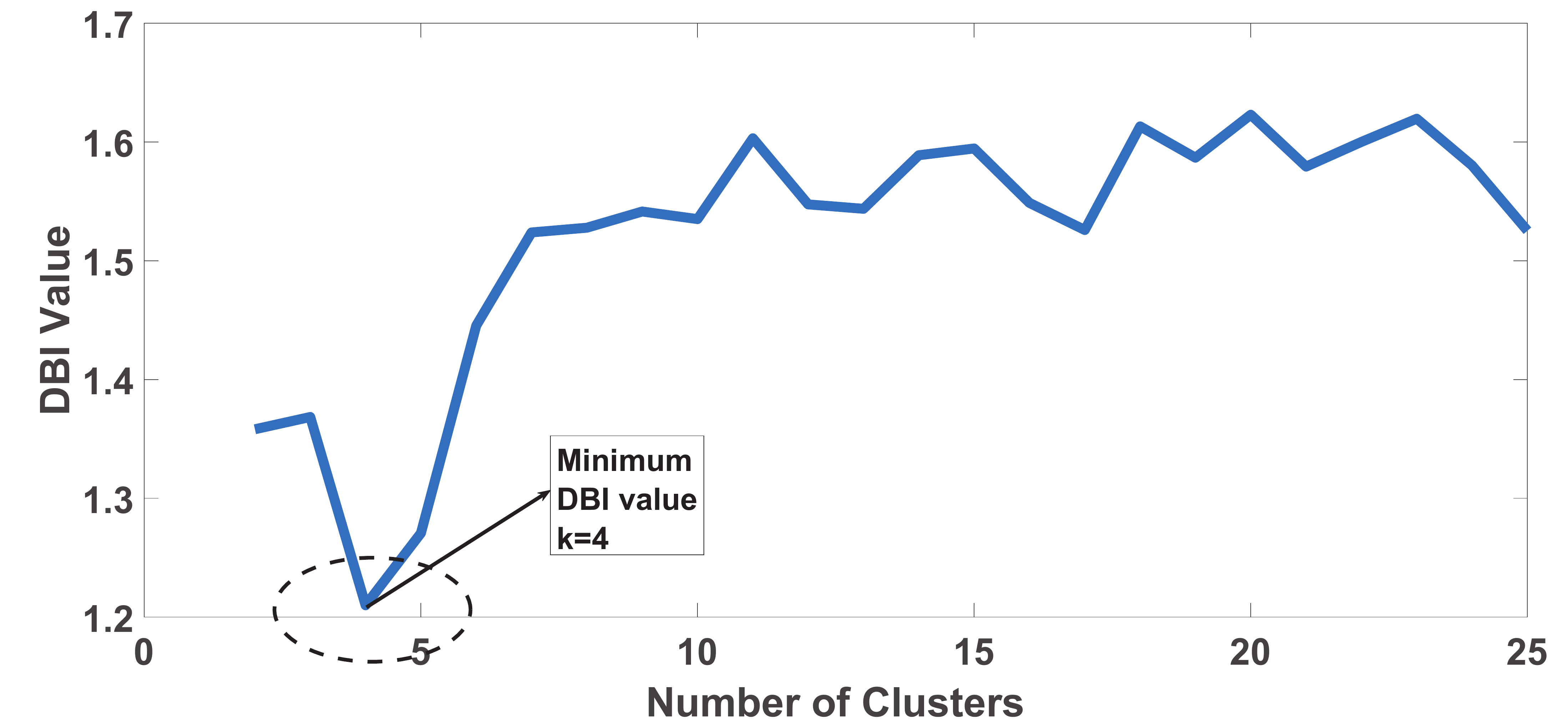}
\caption{Cluster validation index performance for commercial customers.}
\label{fig:DBI}
\end{figure}
Instead of using a single $\alpha$, we calculate a local $\alpha_i$ for each vertex $V_i$ that allows self-tuning of the point-to-point distances, as $\alpha_i = ||V_i-V_K||$, where $V_K$ is the $K$'th neighbor of vertex $V_i$.

\item \textbf{Step II:}
Based on the local scaling parameter $\alpha_i$, the weighted adjacency matrix of the graph $W = (w_{i,j})_{i,j=1,...,n}$ is developed. We have adopted the Gaussian kernel function to build the adjacency matrix $W$ as follows:
\begin{equation}
\label{eq:regression}
w_{i,j} = exp(\frac{-||V_i-V_j||^2}{\alpha_i\alpha_j})
\end{equation}
\item \textbf{Step III:}
After the weighted adjacency matrix is built, SC converts the clustering process to a graph partitioning problem, which divides a graph into $k$ disjoint sets of vertices by removing edges connecting each two groups. When the edges between different sets have low weight and the edges within a set have high weight, a satisfactory partition of the graph is obtained \cite{vonLuxburg2007}. Hence, the objective function of graph partitioning is to maximize both the dissimilarity between the different clusters and the total similarity within each cluster \cite{ISD2007}:
\begin{equation}
\label{eq:ncut1}
N(G) = \min_{A_1,...,A_\eta}\sum_{i=1}^{\eta}\frac{c(A_i,\overline{A_i})}{d(A_i)}
\end{equation}
where, $\eta$ is the number of vertices, $A_i$ is a subset belonging to $V$, $c(A_i,\overline{A_i})$ is the sum of the weights between vertices in $A_i$ and vertices in the rest of the subsets, $d(A_i)$ is the sum of the weights of vertices in $A_i$. It was proved in \cite{NMW2002} that the minimum of $N(G)$ is obtained at the second smallest eigenvector of the Laplacian matrix. Graph Laplacian matrix is the main element of the SC algorithm and constructed using the adjacency matrix $W$ and a diagonal matrix $D$ whose $(i,i)$'th element is the sum of $W$'s $i$'th row. The normalized graph Laplacian is given by \cite{lap1997}:
\begin{equation}
\label{eq:regression}
L = D^{-\frac{1}{2}}WD^{-\frac{1}{2}}
\end{equation}
\item \textbf{Step IV:}
When the associated Laplacian matrix $L$ is constructed, the optimal number of clusters, $k$, needs to be determined to find the best partitioning. This is done using the Davies-Bouldin validation index (DBI) \cite{FMA2015}. The SC was applied to the AMI dataset with different $k$ values and corresponding DBI values for each $k$ were recorded. The value of $k$ for which DBI is minimized is chosen as the optimal number of clusters \cite{FMA2015}. This is shown in Fig. \ref{fig:DBI} for weekday commercial customer data subset. This process is applied to the rest of the data subsets to determine the number of typical load profiles.

\item \textbf{Step V:}
After the optimal $k$ value is obtained using DBI, we compute the first $k$ eigenvectors $x_1,x_2,...,x_k$ of Laplacian matrix $L$ and form the new matrix $X = [x_1,x_2,...,x_k] \in\mathbb{R}^{n\times k}$. Hence, the original data points are mapped to a k-dimensional representation based on these eigenvectors. At this step, we use the K-means algorithm to obtain the $k$ corresponding clusters for the original vertices, $V_i$.
\end{itemize}

\begin{figure*}[hbtp]
      \centering
      \includegraphics[width=2\columnwidth]{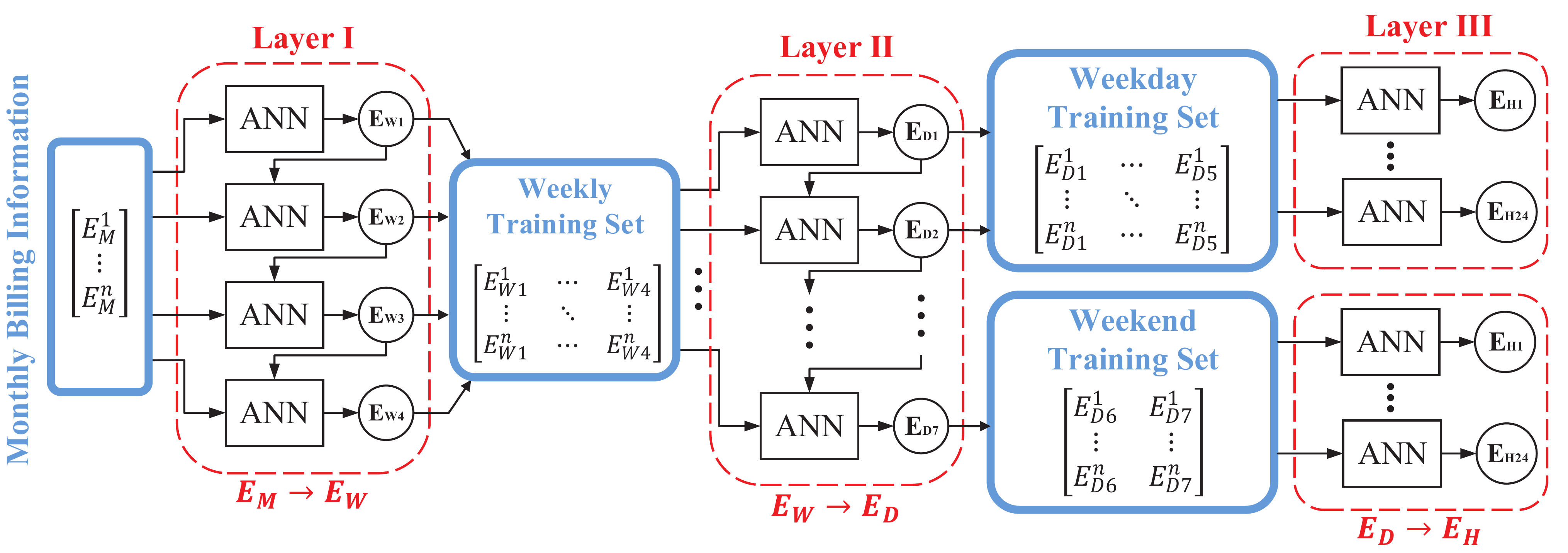}
\caption{Multi-timescale learning structure.}
\label{fig:overall}
\end{figure*}



    
    
    
    
    

\section{Inference of Hourly Energy Consumption}\label{learning}

A MTSL method is assigned and trained for each typical load profile using the available data in the pattern bank defined in Section \ref{cluster}, to map monthly consumption data to hourly load for customers belonging to each class. While hourly load variations cannot be directly observed at the monthly level, a multi-layer structure, where each layer corresponds to the total consumption at different timescales, is able to make this connection between monthly and hourly data with good accuracy. Hence, the MTSL is constructed in a way to keep a high correlation level between inputs-outputs of different layers to maintain layer-wise estimation accuracy. In order to identify variables with high correlation coefficient levels to design the structure of the MTSL, a basic statistical analysis was performed on the AMI dataset, as shown in Table. \ref{table:1.1}. The consumption levels at different timescales are defined as, monthly consumption $E_M$, weekly consumption $E_W$, weekday consumption $E_{D_w}$, weekend consumption $E_{D_{nw}}$, weekday hourly consumption $E_{H_w}$, and weekend hourly consumption $E_{H_{nw}}$, and obtained using hourly SM data history. For different types of customers, the correlation values are shown in Table. \ref{table:1.1} and determined as follows:
\begin{equation}
\label{eq:cor}
\rho(X,Y) = |\frac{\sigma_{X,Y}^2}{\sigma_{X}\sigma_{Y}}|
\end{equation}
where, $X$ and $Y$ are the consumption levels of observed customers at specific timescales, such as monthly or weekly consumption. $\sigma_{X,Y}^2$ is the covariance of $X$ and $Y$, and $\sigma_{X}$ defines the standard deviations of the variable. Using the correlation analysis, a three-layer structure is developed for each type of customer and typical load behavior stored in the pattern bank, as shown in Fig. \ref{fig:overall}. In this figure, Layer I converts total monthly consumption, $E_M$, to the set of weekly consumption values $E_W = \{E_{W1},...,E_{W4}\}$ using ANNs connected in series. To capture the temporal correlation between consumption at consecutive weeks, each week's estimated consumption is also fed to the next ANN corresponding to the following week's consumption. This idea is shown in (\ref{eq:M-W}) and generalized to all the layers of MTSL, as demonstrated in Fig. \ref{fig:overall}:
\begin{table}[htbp]
	\centering
	\caption{Statistical Multi-Timescale Consumption Analysis.}
	\includegraphics[width=0.5\textwidth]{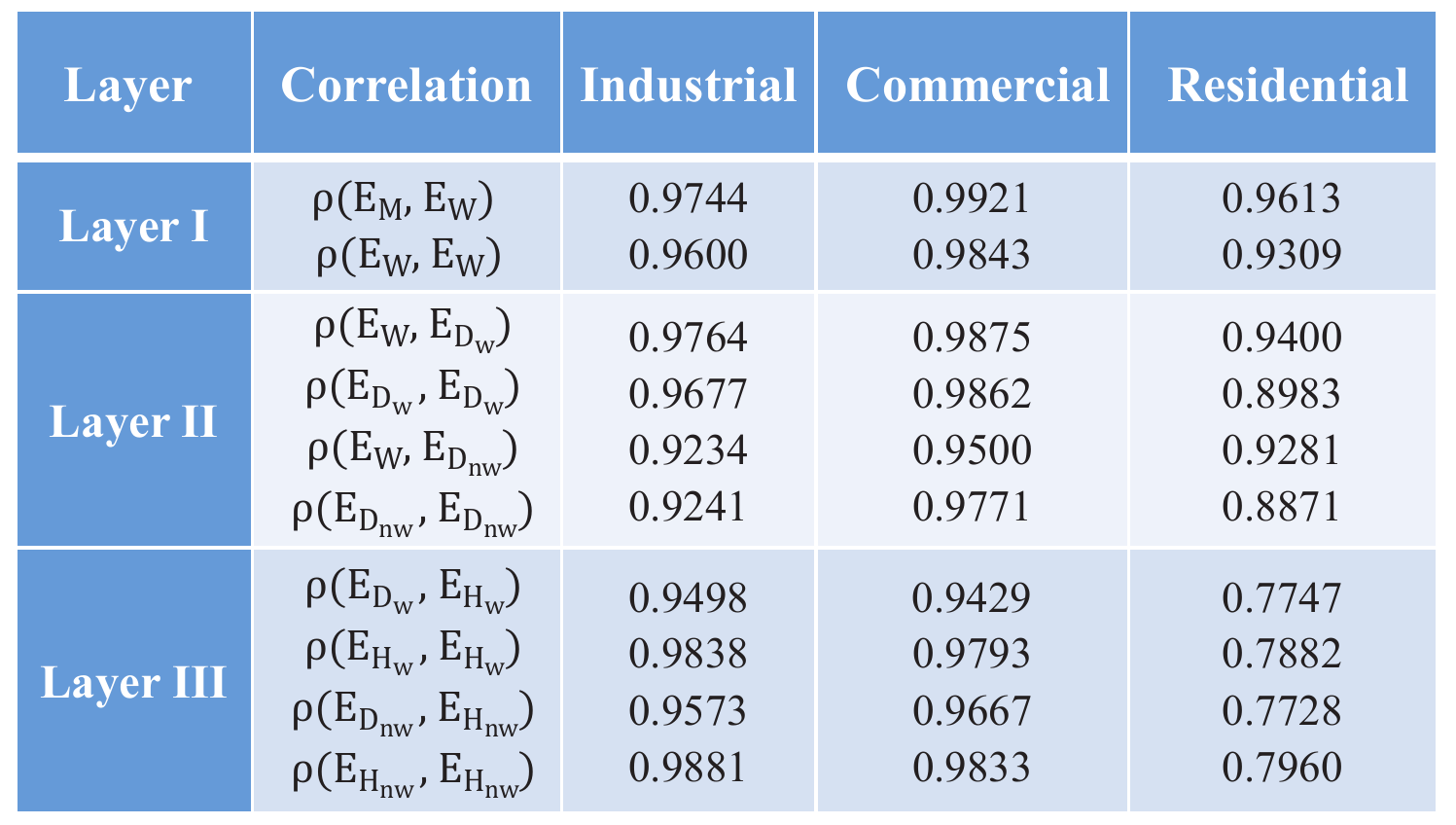}
	\label{table:1.1}
\end{table}
\begin{equation}
\label{eq:M-W}
E_{Wi} = ANN(E_{M},E_{W(i-1)})
\end{equation}

The output of Layer I forms the weekly training set that becomes the input of Layer II. This layer converts weekly consumption, $E_W$, to the set of daily consumption $E_D = \{E_{D1},...,E_{D7}\}$ by various ANNs. Based on the distinct customer behavior on weekdays and weekends, Layer III is trained to map the total daily consumption to hourly consumption $E_H = \{E_{H1},...,E_{H24}\}$. 

At each layer, the dataset is randomly divided into three separate subsets for training (70\% of the total data), validation (15\% of the total data), and testing (15\% of the total data). As a multi-layer structure with a high number of learning parameters, the \textit{overfitting} problem poses a critical risk against reliability of the learned model. Overfitting is a result of model over-flexibility which occurs when the model shows low bias but high variance \cite{overfit2017}. In order to overcome this problem, we have adopted two approaches in this paper: 1) \textit{early stopping mechanism,} in which the training process is terminated as soon as the validation error starts to increase \cite{earlystop2004}. 2) \textit{noise injection,} which improves the robustness of ANNs by injecting small noise to the AMI training sets \cite{Goodfellow2016}.

\section{Proposed Method for Pattern Identification}\label{identify}
\begin{figure}[htbp]
\centering
\subfloat [Industrial (red), commercial (blue), and residential (black) weekday typical load pattern]{
\includegraphics[width=0.45\textwidth]{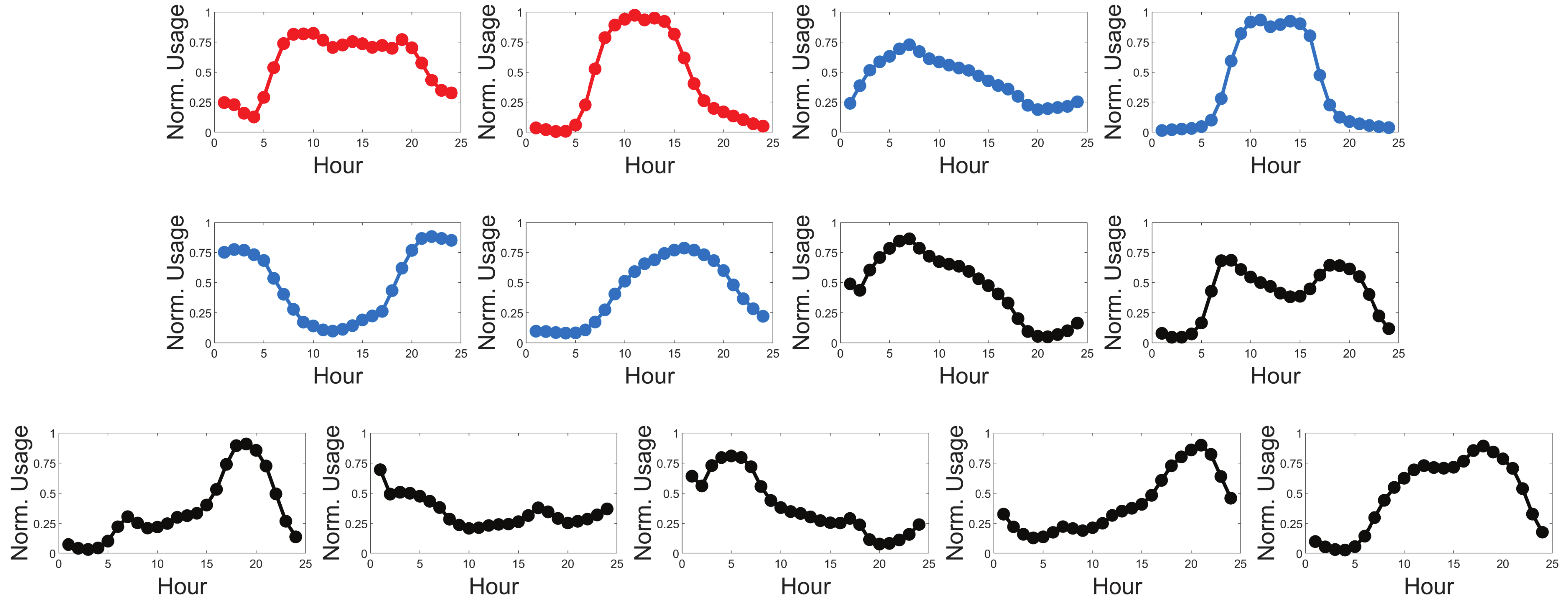}
}
\hfill
\subfloat [Industrial (red), commercial (blue), and residential (black) weekend typical load pattern]{
\includegraphics[width=0.45\textwidth]{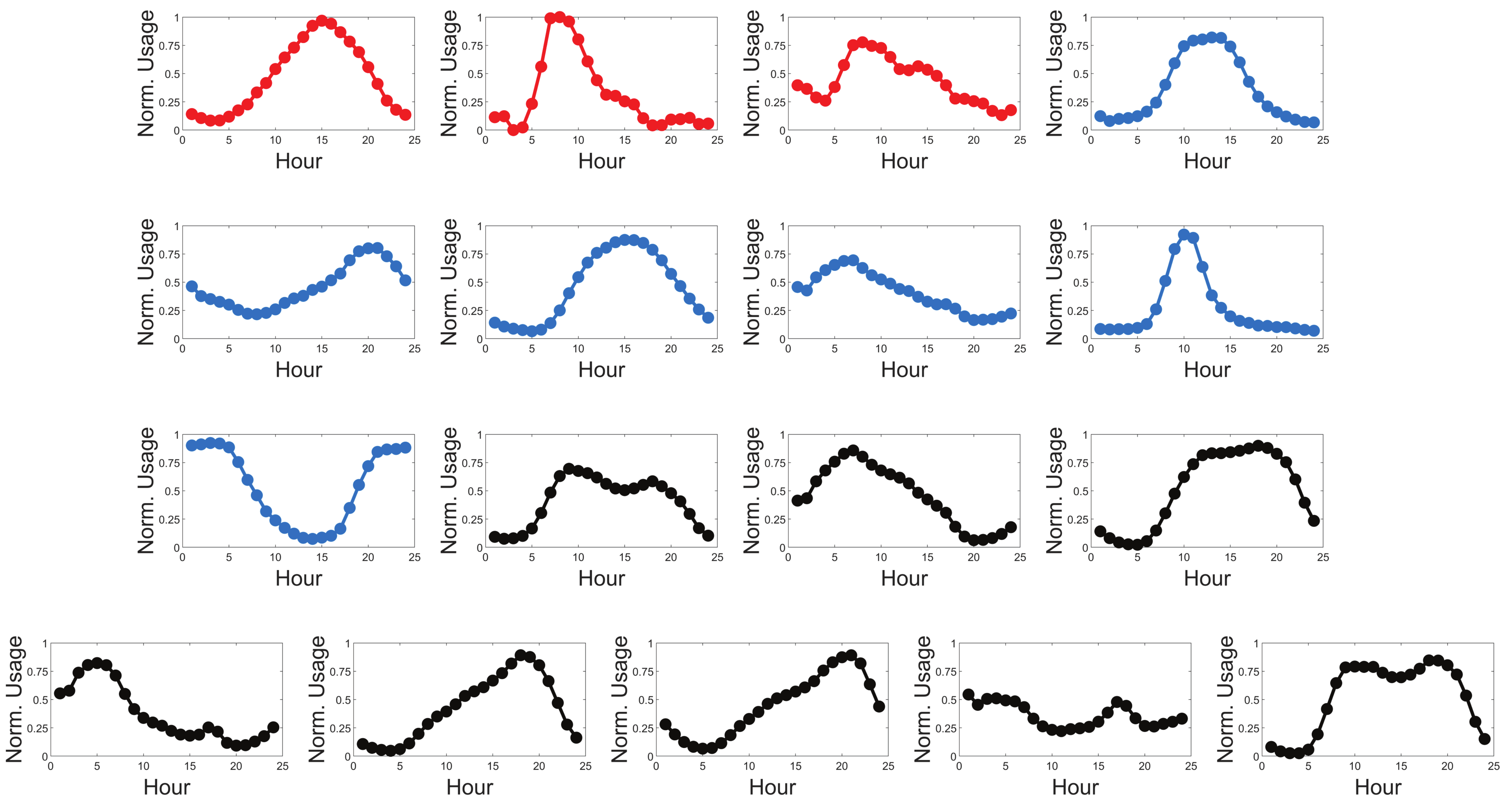}
}
\caption{Consumption pattern bank for industrial, commercial, and residential customers on weekday and weekend.}
\label{fig:cluster}
\end{figure}
In the proposed approach, various MSTL models are assigned to typical consumption patterns. In practice, monthly billing data alone is not enough to determine the typical load profiles of unobserved customers. The pervasive real-time data source in distribution systems is a limited number of feeder-level measurements, such as SCADA voltage and current measurements. In order to identify and allocate the corresponding daily pattern and related MSTL to unobserved customers using only feeder-level measurements, a BCSE-aided RBL method is proposed \cite{RS2010}. This learning algorithm computes the probability of each typical load pattern for an unobserved customer using the residuals of a BCSE algorithm \cite{MEB1995}. Based on the probability values, the most probable class is chosen as the correct underlying profile for unobserved customer. 
\subsection{BCSE}
A BCSE algorithm is tailored for real-time monitoring of distribution systems \cite{MEB1995} \cite{Wang2004}. Compared to traditional state estimation methods that use node voltages as system states, BCSE is shown to improve the computational efficiency and memory requirements by adopting branch currents as state variables. In general, the Weighted Least Square (WLS) algorithm is widely-used to solve the BCSE problem to obtain an estimation of system nodes \cite{AAbur2004}. The objective function of WLS is defined as follows:
\begin{equation}
\label{eq:bcse0}
\min_x J = (z-h(x))^TW(z-h(x))
\end{equation}
where, $z$ is the measurement vector, $x$ is the state vector, i.e., $x = [I_{r},I_{x}]$ with $I_r$ and $I_x$ representing the branch currents' real part and branch currents' imaginary part, $h$ is the nonlinear measurement function associated with measurement $z$, and $W$ denotes the weight matrix that represents the accuracy of measurements. The Gauss-Newton method is adopted to solve this non-convex optimization problem \cite{MEB1995}. The basic idea of Gauss-Newton method is to find a solution for $\nabla_x J=0$, where $\nabla_x J$ denotes the gradient of $J$ with respect to state variables. The iterative processes of the algorithm are as follows:
\begin{equation}
\label{eq:bcse3}
G(x) = H^T(x)WH(x)
\end{equation}
\begin{equation}
\label{eq:bcse4}
[G(x^m)]\Delta x^m = H^T(x^m)W(z-h(x^m))
\end{equation}
\begin{equation}
\label{eq:bcse5}
x^{m+1} = x^{m} + \Delta x^{m}
\end{equation}
where, $H$ is the Jacobian matirx of the measurement function $h(x)$, $G$ is the gain matrix, and $m$ is the iteration number. 

\subsection{Load Pattern Assignment by RBL}

To identify the underlying daily consumption pattern for unobserved customers, the following steps are performed:
\begin{itemize}
\item \textbf{Stage I:} Select a class, denoted as $i$, from the daily consumption pattern bank, for unobserved customer $j$.
\item \textbf{Stage II:} Use the MSTL of the selected class to generate hourly pseudo load values from the customer's monthly billing data.
\item \textbf{Stage III:} Run the BCSE using the generated pseudo load values. Observe the residuals. The residuals of each estimator can be obtained by comparing the real measurements with estimated values. 
\item \textbf{Stage IV:} Define probability $p_{i,j}$ as: ``the probability that class $i$ is the correct average daily consumption profile for customer $j$.'' Applying the Bayes theorem and assuming a Gaussian distribution for measurement error, a recursive expression for updating this probability over time is obtained as follows \cite{RS2010} \cite{SJW2001}:
\begin{figure}[htbp]
	\centering
	\includegraphics[width=3.35in]{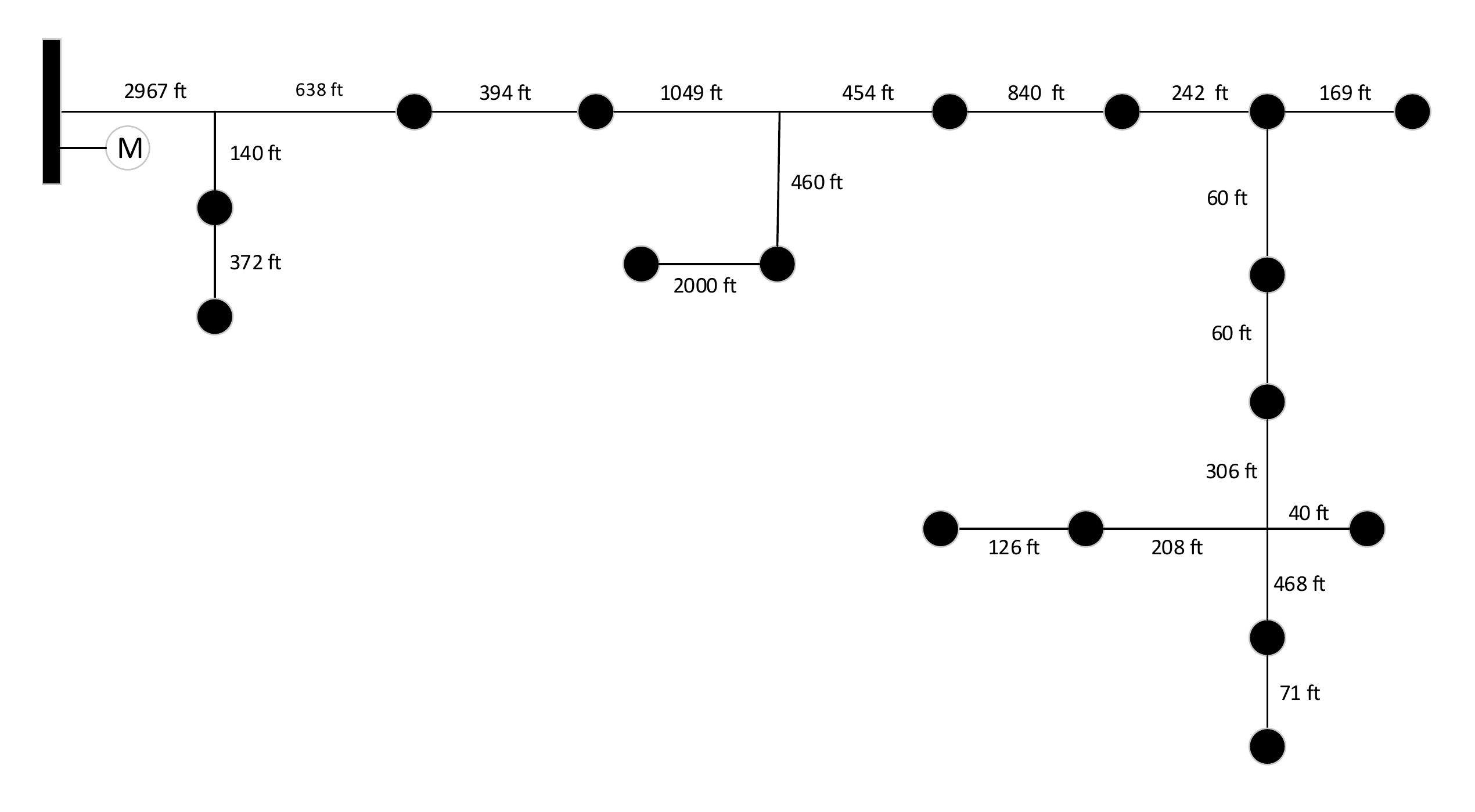}
	\caption{A 18-node real utility feeder case.}
	\label{fig:topology}
\end{figure}
\begin{figure}[htbp]
	\centering
	\includegraphics[width=3.35in]{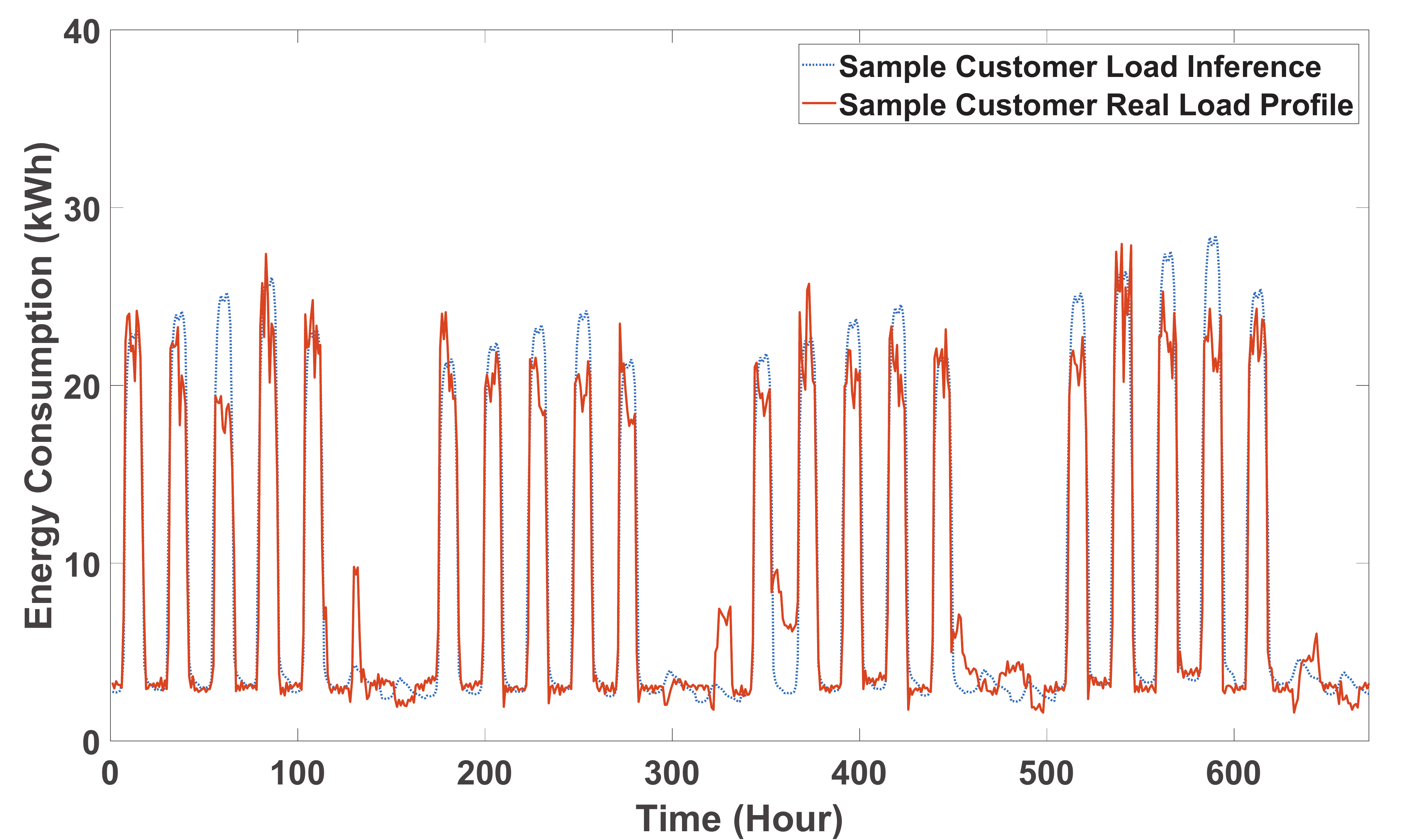}
	\caption{Comparison of hourly load inference with real load profile.}
	\label{fig:sample}
\end{figure}
\begin{equation}
\label{eq:RBL2}
p_{i,j}^o=\frac{\exp(-\frac{1}{2}r_{i,j}^{o^T}\cdot \Phi\cdot r_{i,j}^o)p_{i,j}^{o-1}}{\sum_{t=1}^N \exp(-\frac{1}{2}r_{t,j}^{o^T}\cdot \Phi\cdot r_{t,j}^o)p_{t,j}^{o-1}}
\end{equation}
where, $N$ is the number of MSTL models for the specific customer type, $o$ is the iteration count, $r_{i,j}^o$ is the residual vector of the $i$'th class, $\Phi$ is a diagonal matrix that represents the inverse of the variances corresponding to the residual components, which increases the speed of convergence.
\item \textbf{Stage V:} Go back to Stage II to repeat this process for another candidate typical daily load profile for customer $j$.
\item \textbf{Stage VI:} Identify the underlying daily load profile for the unobserved customer, $i^*$, as the most probable class: $i^*=\argmaxA_i p_i^j$.
\item \textbf{Stage VII:} Repeat the above process for all unobserved customers until the average daily load profiles of all customers are identified.
\item \textbf{Stage VIII:} Perform online BCSE for real-time system monitoring using MTSL-based pseudo hourly load estimations obtained from the assigned classes to unobserved customers.
\end{itemize}

The main advantage of the RBL is exponential rejection of the wrong load patterns and low computational complexity which is advantageous in large distribution systems \cite{RS2010}. 

\section{Numerical Results}\label{result}
\begin{figure}[htbp]
	\centering
	\includegraphics[width=3.35in]{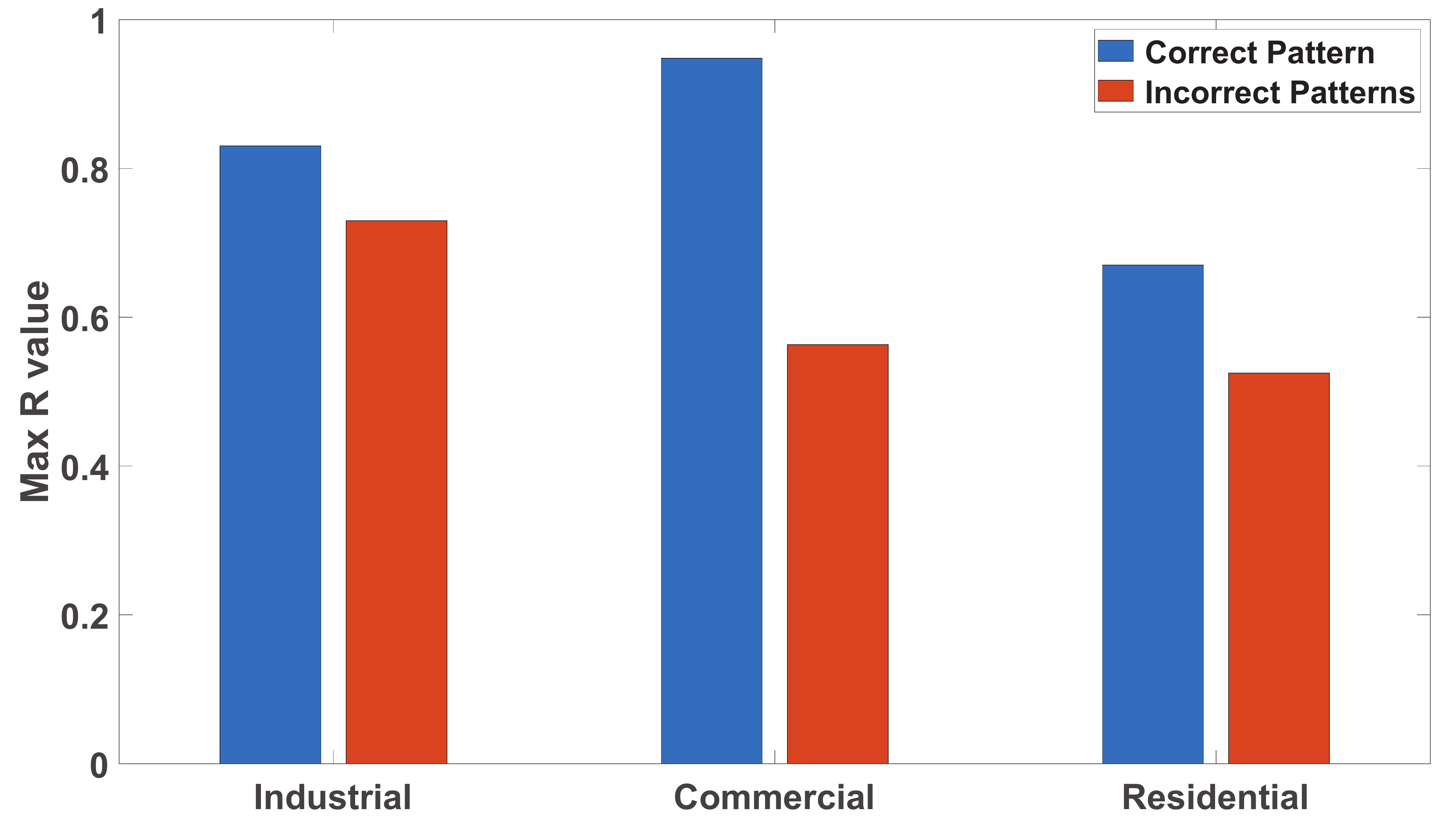}
	\caption{Customer level load estimation result.}
	\label{fig:customer_level}
\end{figure}
The proposed observability enhancement framework is tested for unobserved customers on a real distribution feeder, shown in Fig. \ref{fig:topology}. This feeder contains three types of loads: industrial (3\%), commercial (20\%), and residential (77\%) loads. The proposed method is compared with two existing load estimation approaches adopted from \cite{YR2018} and \cite{DT2015}, in terms of accuracy.

\subsection{SC Algorithm Performance} 
Based on the AMI dataset, the SC algorithm is utilized to classify different load shapes and to create the consumption pattern banks. Fig. \ref{fig:cluster} shows typical load patterns for different types of customers for weekdays and weekends. As shown in Fig. \ref{fig:cluster}, the numbers of typical load profiles in weekdays are normally smaller than that of weekends. Compared to the diverse activities in weekends, customers have relatively few normative load behaviors in weekdays. Also, as expected, the residential customers have more load patterns than industrial and commercial customers due to the higher variation of residential load behaviors.
\subsection{Pseudo Measurement Generation Performance}

After consumption pattern banks have been developed from AMI data of observed systems, the muti-layer MSTL models are trained and tested on the feeder shown in Fig. \ref{fig:topology}. In this case, the test feeder is considered to be a fully unobserved network in which no customer is equipped with SMs. To reduce the error of the learning model, the MTSL method has been tested over 12-month load data. Fig. \ref{fig:sample} shows the comparison between hourly load inference of one sample customer, obtained from monthly billing data, and real load profile during that month. As can be seen, the pseudo hourly load samples are able to accurately track the customer's real consumption. 
\begin{figure}[htbp]
\centering
\subfloat [Sample feeder daily load inference results in weekday]{
\includegraphics[width=3.35in]{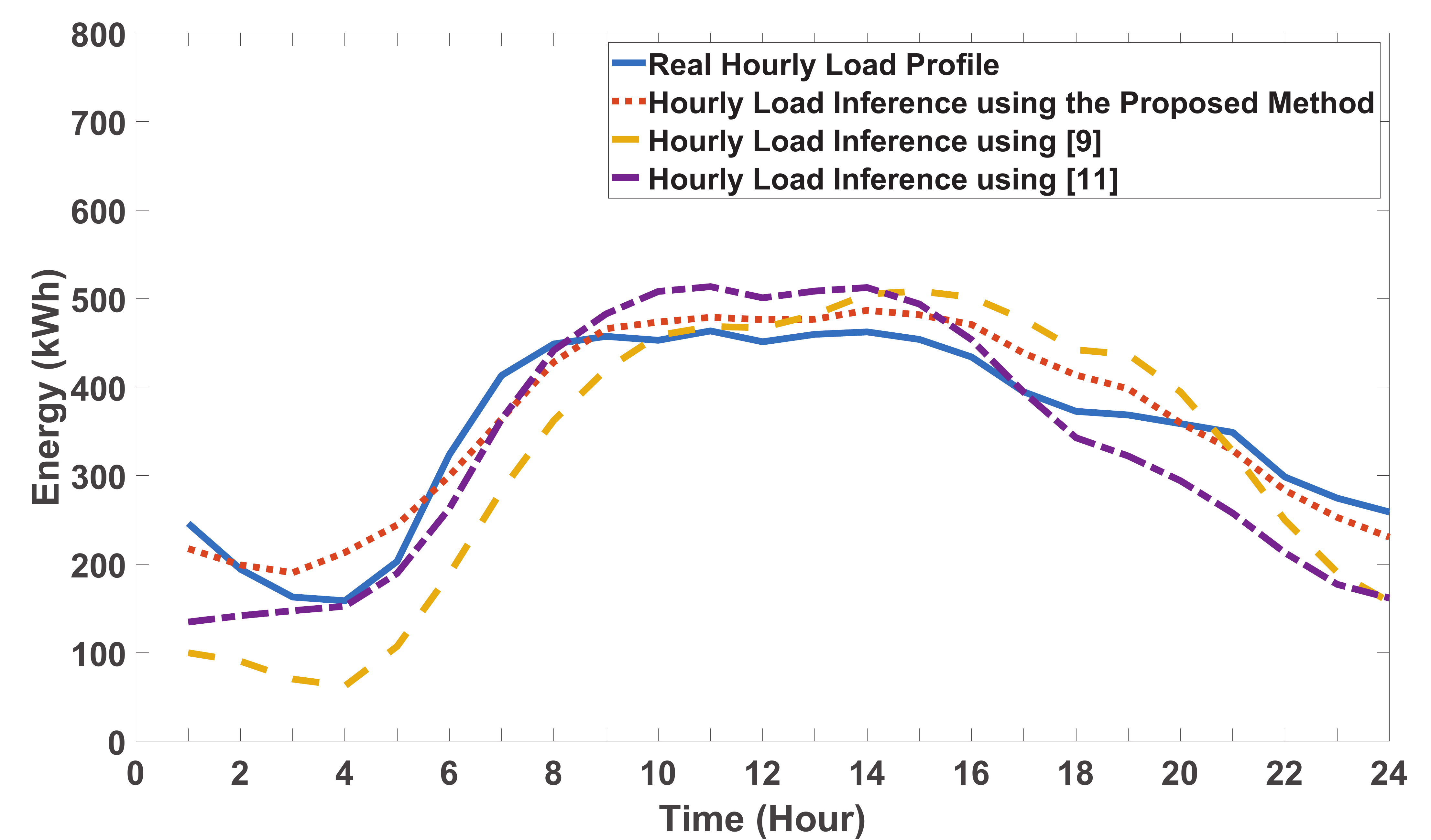}
}
\hfill
\subfloat [Sample feeder daily load inference results in weekend]{
\includegraphics[width=3.35in]{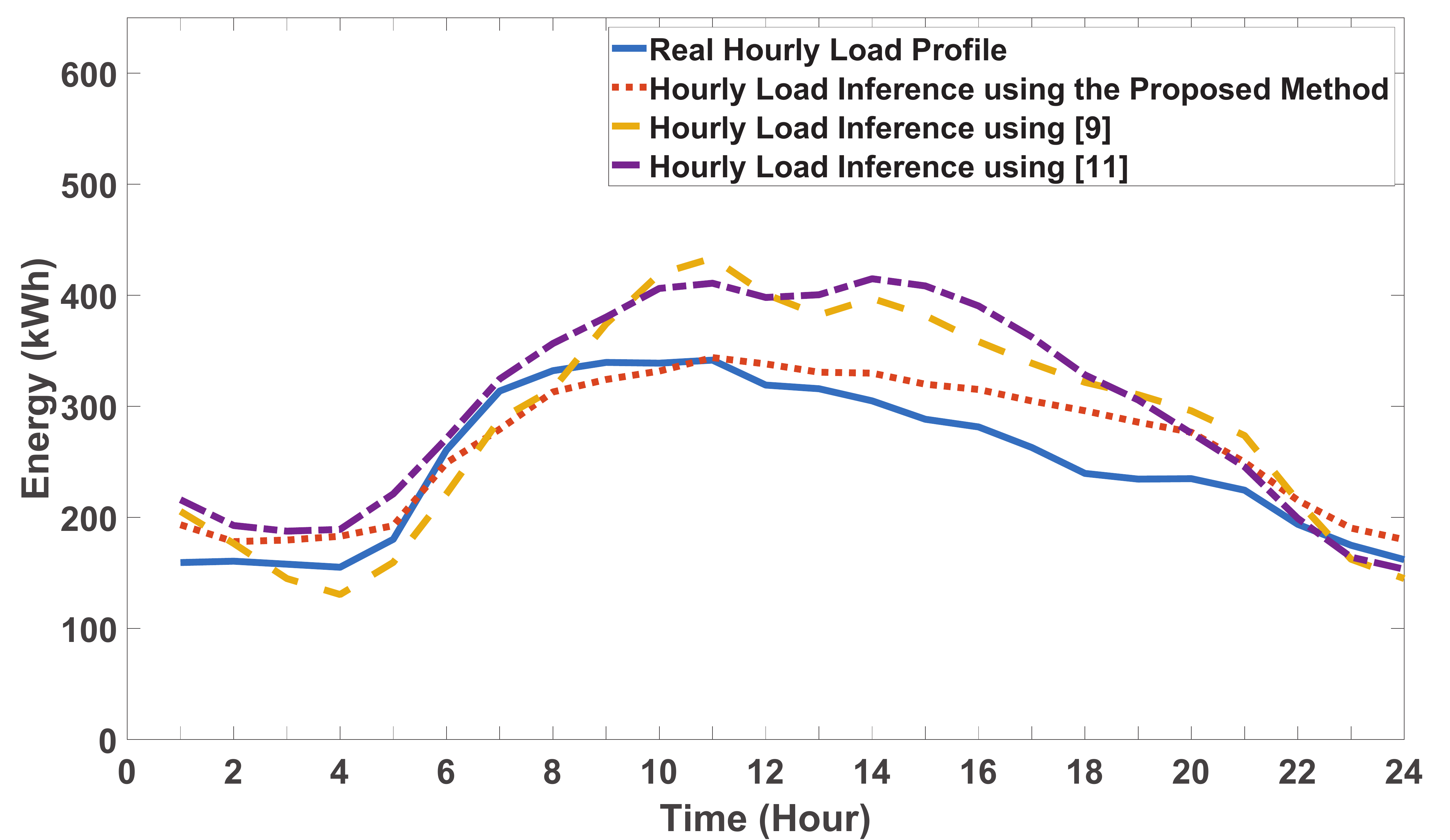}
}
\caption{Comparison of load inference results.}
\label{fig:feeder_level}
\end{figure}
Fig. \ref{fig:customer_level} presents the accuracy comparison of load estimation for different types of customers. The monthly data of test customers are used as the input of all MSTL models. The \textit{goodness-of-fit} measure, $R$, is used to assess the accuracy of the result, with $R = 1$ indicating a perfect fit. The $R$ values are used to measure the accuracy of MTSLs corresponding to correct and incorrect daily pattern consumption classes for all customers. As expected, the MTSL load estimation model corresponding to the correct underlying consumption class for the customers has a better accuracy, compared to the incorrect one. This further supports the correct functionality of RBL, as described in the next subsection. Also, as shown in Fig. \ref{fig:customer_level}, for industrial and commercial customers, the learning model yields more accurate estimations compared to the residential customers due to lower consumption volatility. In contrast, for residential customers, the diversity and complexity of human activities lead to less accurate estimations. 

Fig. \ref{fig:feeder_level} shows the feeder-level load estimation results in weekdays and weekends for our proposed learning model and two existing methods in the literature \cite{YR2018} \cite{DT2015}. The Mean Absolute Percentage Error (MAPE) criterion is utilized to evaluate the accuracy of estimation methods:
\begin{figure}[htbp]
\centering
\subfloat [Sample industrial customer identification]{
\includegraphics[width=3.35in]{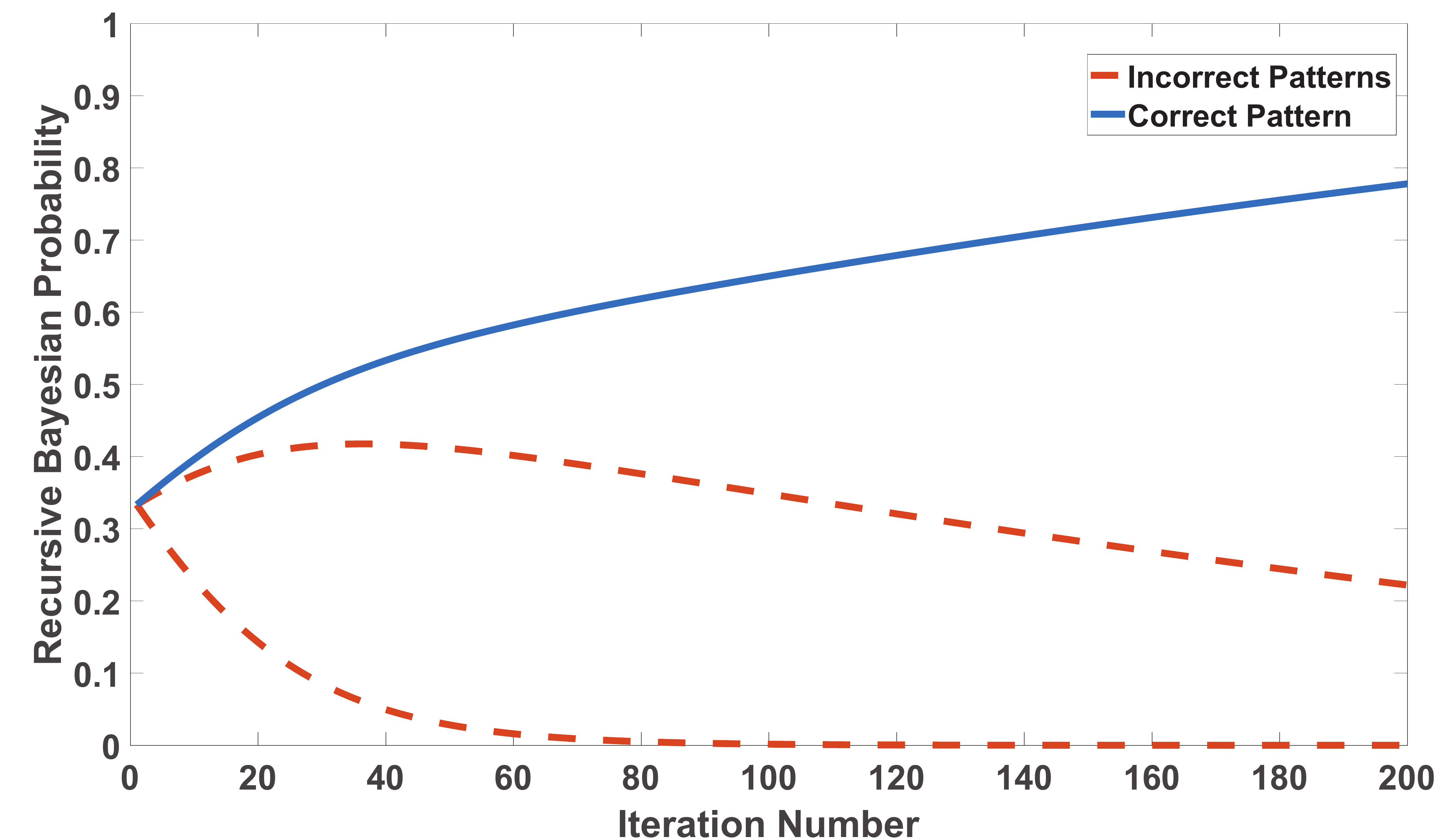}
}
\hfill
\subfloat [Sample commercial customer identification]{
\includegraphics[width=3.35in]{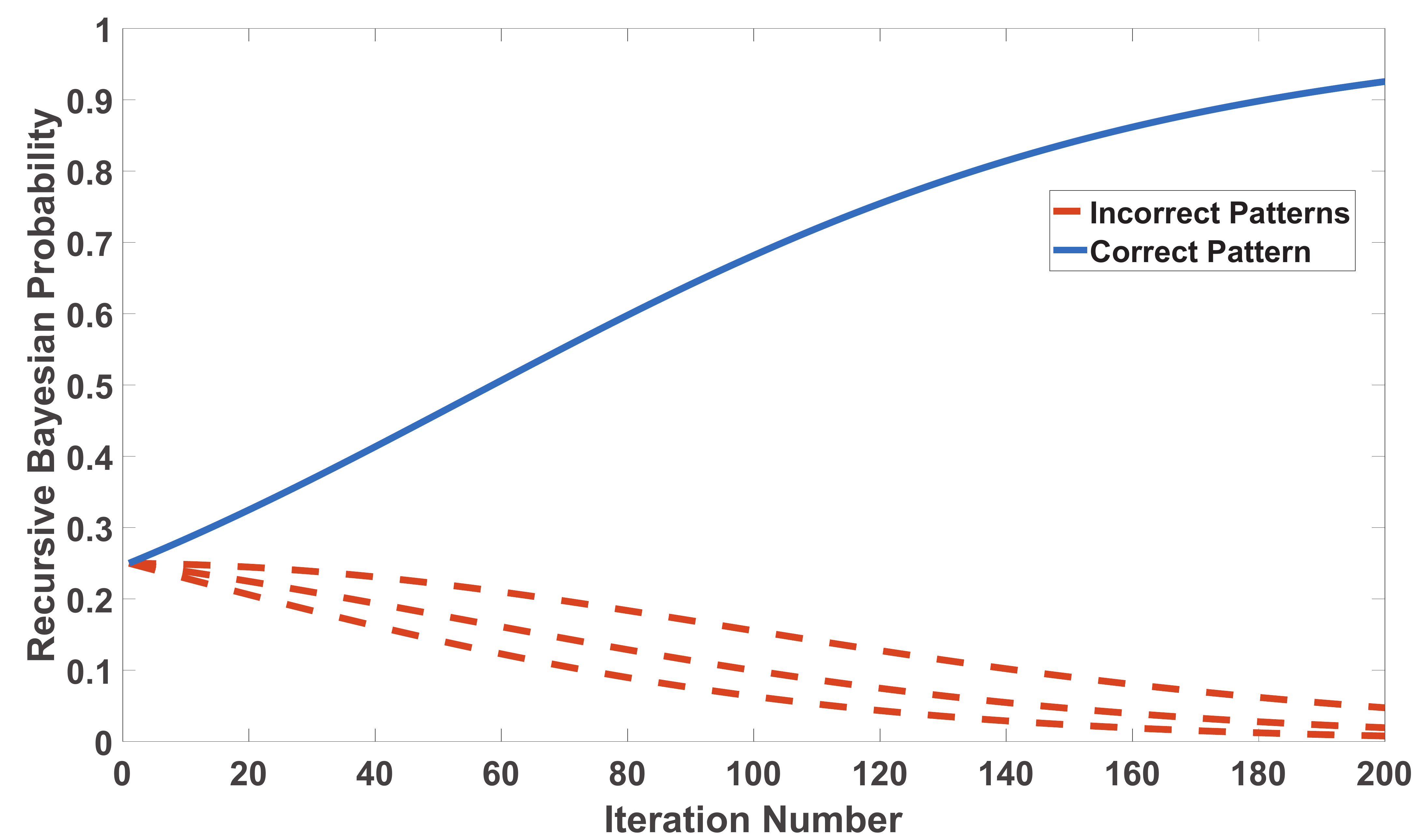}
}
\hfill
\subfloat [Sample residential customer identification]{
\includegraphics[width=3.35in]{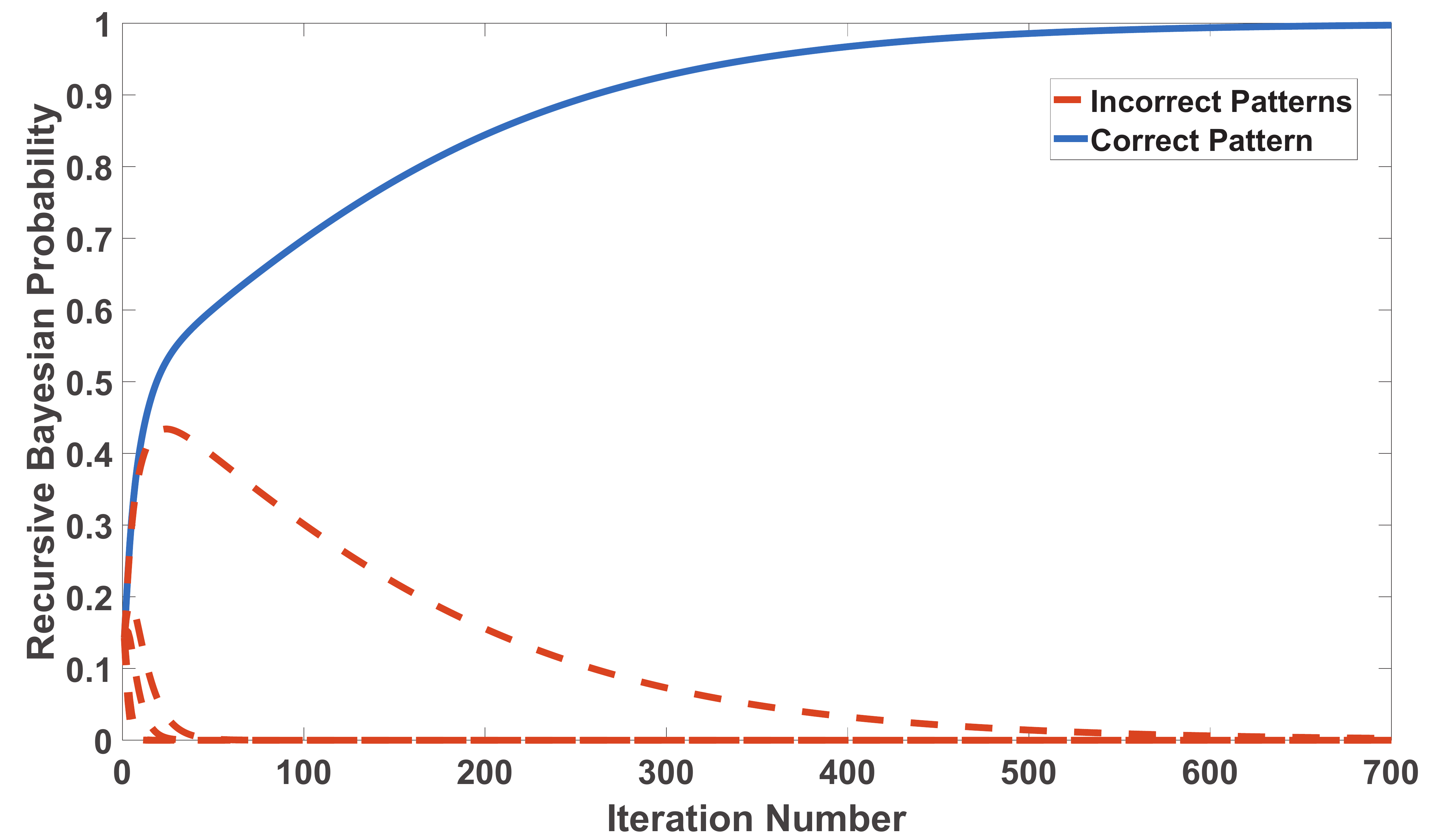}
}
\hfill
\caption{Performance of BCSE-aided RBL daily profile identification method for three types of customers.}
\label{fig:Ihist}
\end{figure}
\begin{equation}
\label{eq:MAPE}
M = \frac{100\%}{n_s}\sum_{t=1}^{n_s}|\frac{A(t)-E\{A(t)\}}{A(t)}|
\end{equation}
where, $A$ is the actual load value and $E\{\cdot\}$ is the mean operator. As is demonstrated in these figures, the estimation MAPE values for the proposed method are $\{7.40\%, 10.02\%\}$ for weekdays and weekends, respectively. On the other hand, the proposed methods in \cite{YR2018} and \cite{DT2015} show average MAPE of \{19.47\%, 20.32\%\} and \{13.79\%, 21.16\%\} over the test set. Hence, based on this AMI dataset and the test feeder, the proposed method shows a better accuracy for hourly load inference compared to the previous works.

\subsection{Load Pattern Identification}
\begin{figure}[htbp]
\centering
\subfloat  [Voltage magnitude component error]{
\includegraphics[width=3.35in]{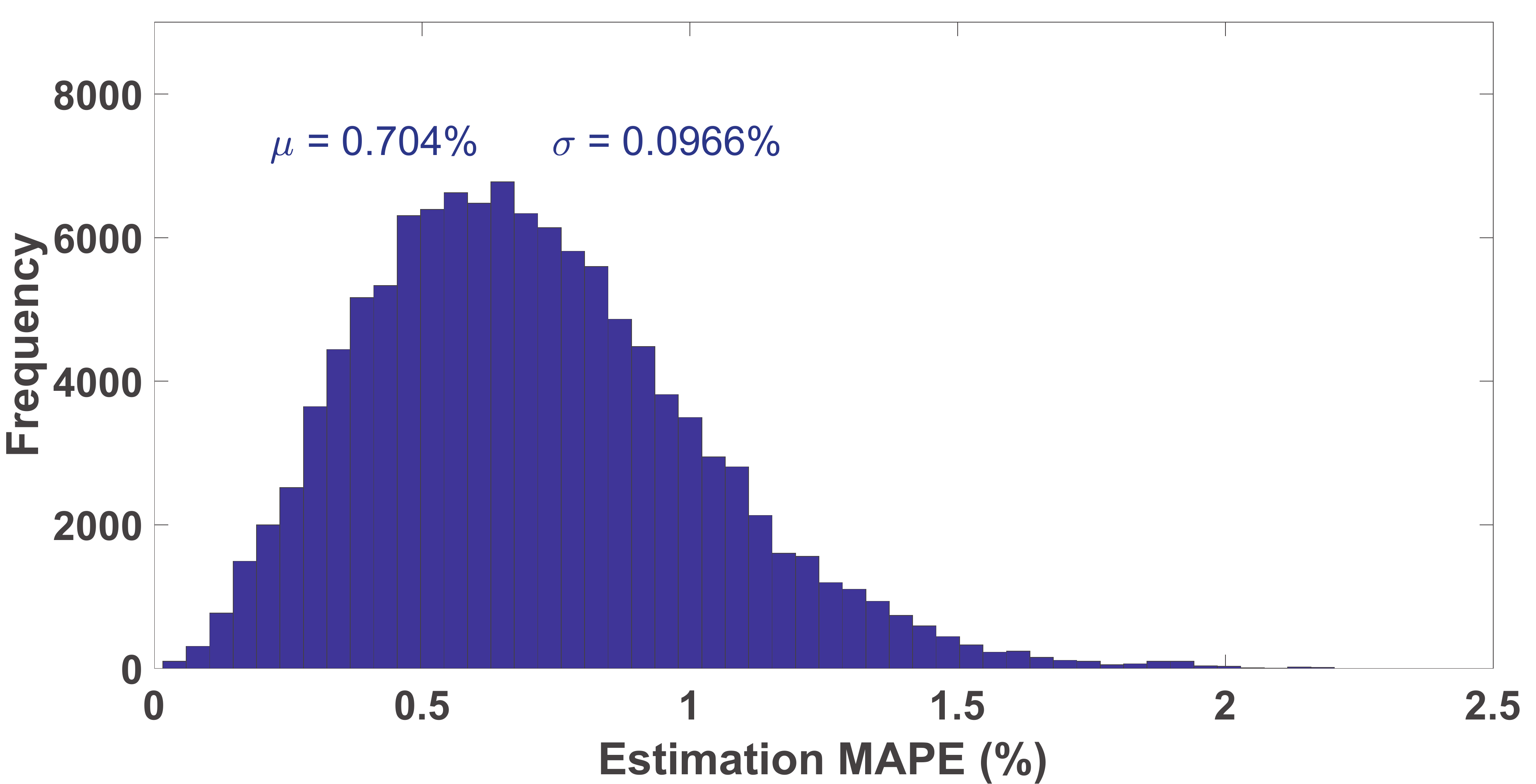}
}
\hfill
\subfloat [Voltage phase component error]{
\includegraphics[width=3.35in]{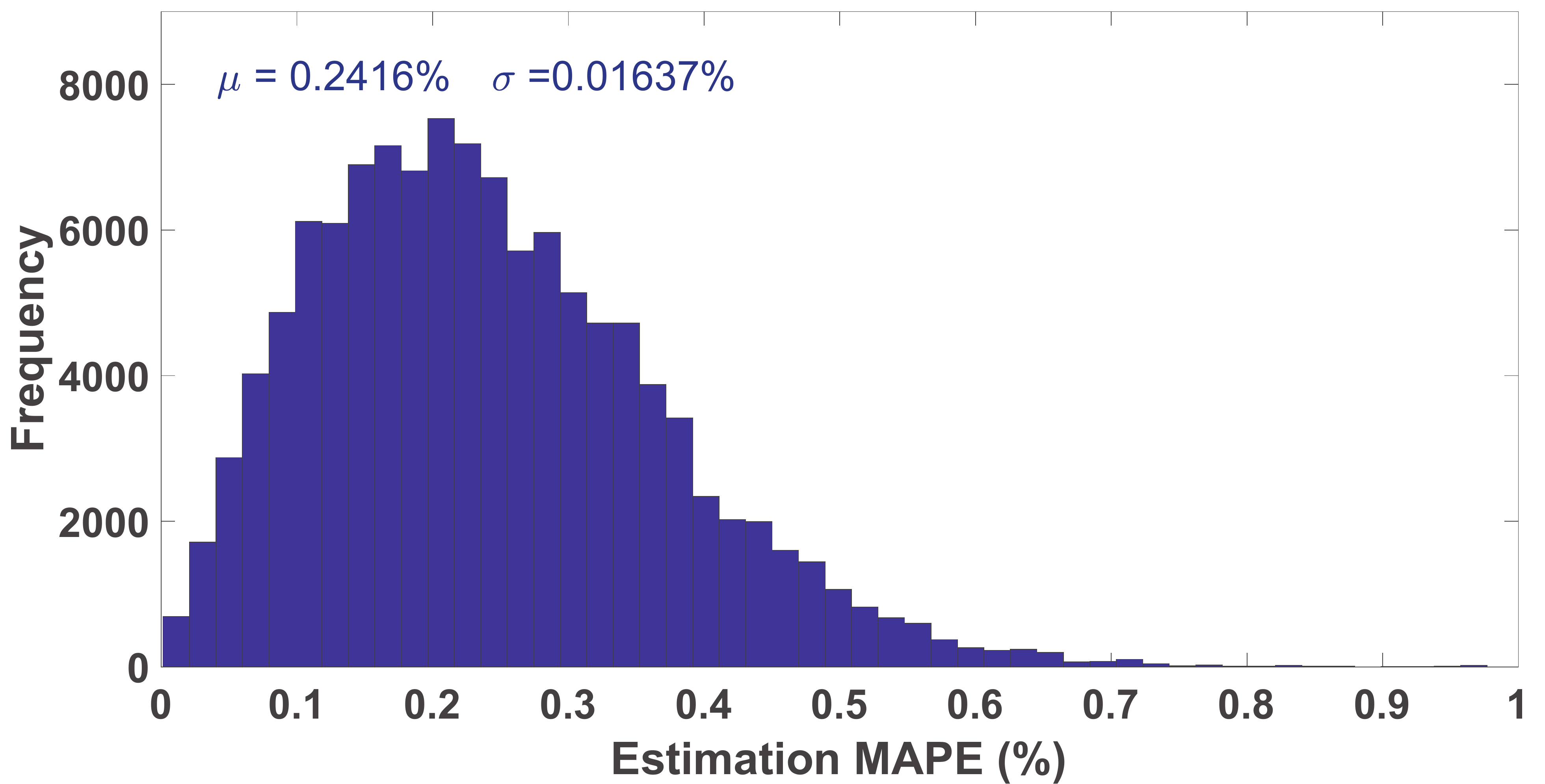}
}
\caption{BCSE-based state estimation performance using the proposed load inference model.}
\label{fig:BCSEerror}
\end{figure}
The performance of the BCSE-aided pattern identification scheme was tested on three cases of different types of customers, corresponding to industrial, commercial, and residential loads. A Phasor Measurement Unit (PMU) was placed at the main bus of the test feeder to provide the real measurement value for BCSE. Pseudo hourly load estimations were extracted from unobserved customers' monthly billing data, for different candidate daily consumption profiles in the databank. According to the residuals, the graphs in Fig. \ref{fig:Ihist} show the probabilities assigned by the RBL algorithm to the correct and incorrect load patterns available in the typical daily load profile bank. For all types of customers, the algorithm effectively identifies the MTSL model corresponding to the correct daily consumption pattern, by assigning the highest probability value to it.  

\subsection{State Estimation Performance}
After hourly pseudo measurement samples are generated for every unobserved customer using the proposed method, BCSE can be performed in real-time over the test feeder given the introduced data-driven redundancy. The error distribution of real-time state estimation is shown in Fig. \ref{fig:BCSEerror} for voltage magnitude and phase components. As is demonstrated in the figure, based on the proposed load estimation approach, BCSE can obtain accurate system state estimation with magnitude and phase angle estimation mean errors of $0.70\%$ and $0.24\%$, respectively. 

\section{Conclusion}\label{conclusion}
In this paper, we have presented a data-driven method for load estimation to improve the observability of distribution systems without AMI. The proposed method is able to extract hourly load estimations from monthly billing data for all types of customers, including residential, commercial, and industrial. Moreover, this approach can identify the average daily load pattern of unobserved customers using a BCSE-aided probabilistic learning method. The proposed method is successfully validated on a real utility feeder with real SM data and has been able to improve the performances of existing methods in the literature. 

\ifCLASSOPTIONcaptionsoff
  \newpage
\fi



\bibliographystyle{IEEEtran}
\bibliography{IEEEabrv,./bibtex/bib/IEEEexample}
\end{document}